\documentclass[pra,aps,twocolumn,floatfix,showpacs]{revtex4}
\usepackage{graphicx,amsmath,amssymb,times}

\begin{document}
\title{Momentum distribution of the insulating phases of the extended Bose-Hubbard model}
\author{M. Iskin$^1$ and J. K. Freericks$^2$}
\affiliation{$^1$Joint Quantum Institute, National Institute of Standards and Technology, and 
University of Maryland, Gaithersburg, MD 20899-8423, USA. \\
$^2$Department of Physics, Georgetown University, Washington, DC 20057, USA.}
\date{\today}

\begin{abstract}
We develop two methods to calculate the momentum distribution of the
insulating (Mott and charge-density-wave) phases
of the extended Bose-Hubbard model with on-site and 
nearest-neighbor boson-boson repulsions on $d$-dimensional hypercubic 
lattices. First we construct the random phase approximation result, 
which corresponds to the exact solution for the infinite-dimensional 
limit. Then we perform a power-series expansion in the hopping $t$ 
via strong-coupling perturbation theory, to evaluate the momentum distribution 
in two and three dimensions; we also use the strong-coupling theory to 
verify the random phase approximation solution in infinite dimensions. Finally, we briefly 
discuss possible implications of our results in the context of ultracold 
dipolar Bose gases with dipole-dipole interactions loaded into optical lattices.
\end{abstract}

\pacs{03.75.Lm, 37.10.Jk, 67.85.-d}
\maketitle

\section{Introduction}
\label{sec:introduction}

Ultracold atomic gases loaded into optical lattices have been proven
to be ideal systems for studying Hubbard-type Hamiltonians~\cite{jaksch}, 
the most successful of which has been the Bose-Hubbard (BH) model. 
This model has three terms~\cite{fisher}: 
a kinetic energy term which allows for the tunneling of the bosons 
between nearest-neighbor lattice sites, a potential energy term which 
is given by the repulsion between bosons that occupy the same 
lattice site, and a chemical potential term which fixes the 
number of bosons. The phase diagram of this model has 
been known for a long time~\cite{fisher, freericks-1, freericks-2, stoof, prokofiev-1, prokofiev-2}. 
The competition between the kinetic and potential energy terms
leads to two phases: a Mott insulator (Mott) when the kinetic energy 
is much smaller than the potential energy  
and a superfluid otherwise. The Mott phase has an excitation gap 
and is incompressible, and therefore, the bosons are localized and incoherent, 
so that a slight change in the chemical potential does not change the number 
of bosons on a particular lattice site. The superfluid phase, however, 
is gapless and compressible, and the bosons are delocalized and move 
coherently. Both of these phases, as well as the transition between the 
two, have been successfully observed with ultracold point-like Bose 
gases loaded into optical lattices~\cite{greiner, spielman-1, spielman-2, bloch}. 

The on-site BH model takes only the on-site boson-boson repulsion into 
account, \textit{i.e.} the interaction is short-ranged. A more 
general extended BH model is required when longer-ranged interactions 
are not negligible, \textit{e.g.} Coulomb or dipole-dipole interactions. 
For instance, an ultracold dipolar Bose gas can be realized in many 
ways with optical lattices~\cite{goral}: (ground-state) heteronuclear 
molecules which have permanent electric dipole moments, 
Rydberg atoms which have very large induced electric dipole moment, 
or Chromium-like atoms which have large intrinsic magnetic moment, 
etc. can be used to generate sufficiently strong long-ranged 
dipole-dipole interactions.
The qualitative phase diagram of this model has also been known for a 
long time~\cite{bruder, parhat, otterlo, kuhner, kovrizhin, menotti, iskin}, and it has two
additional phases: a charge-density wave (CDW) as shown in Fig.~\ref{fig:mf} 
and a supersolid. Similar to the Mott phase, the CDW phase is an 
insulator with an excitation gap 
and it is incompressible. The main difference is that an integer number of 
bosons occupy every lattice site in the Mott phase, while the CDW 
phase has a crystalline order in the form of staggered boson numbers 
(different occupancy on different sublattices).
As the name suggests, a supersolid phase~\cite{leggett}, however, has both the superfluid 
and crystalline orders, \textit{i.~e.} both CDW and superfluid phases coexist. 
There is some evidence that this phase exists only in dimensions higher 
than one~\cite{otterlo, kuhner}. 

There has been experimental progress in constructing ultracold dipolar 
gases of molecules, namely ground-state K-Rb molecules, from a mixture of fermionic $^{40}$K and 
bosonic $^{87}$Rb atoms~\cite{ye-1,ye-2}. 
While this K-Rb is a fermionic molecule, similar principles will allow one to 
also create bosonic dipolar molecules by simply changing the atomic isotopes. 
Motivated by these achievements, in this paper, we analyze the 
momentum distribution of the insulating phases of the extended BH model, 
which is the most common probing technique used in atomic systems 
to identify different phases.

The remainder of this paper is organized as follows. After introducing the
model Hamiltonian in Sec.~\ref{sec:ebh}, we develop two methods 
in Sec.~\ref{sec:md} to calculate the momentum distribution of the 
insulating (Mott or charge-density-wave) phases of the extended Bose-Hubbard model. 
First we use the random phase approximation (RPA) in Sec.~\ref{sec:rpa}, and
then we perform a power series expansion in the hopping $t$ via the 
strong-coupling perturbation theory in Sec.~\ref{sec:sc}. 
The numerical analysis of the momentum distribution obtained from these 
methods are discussed in Sec.~\ref{sec:numerics}, and a brief 
summary of our conclusions is presented in Sec.~\ref{sec:conclusions}.
Finally in Appendix~\ref{sec:app}, we comment on some of the issues 
regarding the Wannier functions in the CDW phase.

\section{Extended Bose-Hubbard Model}
\label{sec:ebh}

We consider the following extended BH Hamiltonian with on-site and 
nearest-neighbor boson-boson repulsions
\begin{align}
\label{eqn:ebhh}
H = &- \sum_{i,j } t_{ij} b_i^\dagger b_j 
+ \frac{U}{2} \sum_i \widehat{n}_i (\widehat{n}_i-1) \nonumber \\
&+ \sum_{ i,j } V_{ij}\widehat{n}_i \widehat{n}_j -\mu \sum_i \widehat{n}_i,
\end{align}
where $t_{ij}$ is the tunneling (or hopping) matrix between sites $i$ and $j$, 
$b_i^\dagger$ ($b_i$) is the boson creation (annihilation) operator at site $i$,
$\widehat{n}_i = b_i^\dagger b_i$ is the boson number operator, 
$U>0$ is the strength of the on-site and 
$V_{ij}$ is the longer-ranged boson-boson repulsion between bosons at sites $i$ and $j$, 
and $\mu$ is the chemical potential.
In this paper, we assume $t_{ij}$ is a real symmetric matrix with elements
$t_{ij} = t$ for $i$ and $j$ nearest neighbors and $0$ otherwise and similarly 
for $V_{ij}$ (equal to $V>0$ for $i$ and $j$ nearest neighbors and zero otherwise), 
and consider a $d$-dimensional hypercubic lattice with $M$ sites.
Note that we work on a periodic lattice without an external trapping potential.
We also assume $U > z V$ where $z=2d$ is the lattice coordination number 
(number of nearest neighbors). 

The ground-state phase diagram of this model Hamiltonian has been 
studied extensively in the literature including 
the mean-field~\cite{bruder}, quantum Monte Carlo~\cite{parhat, otterlo},
density-matrix renormalization group~\cite{kuhner},  
Gutzwiller ansatz~\cite{kovrizhin, menotti}, 
and strong-coupling expansion and scaling theory~\cite{iskin} techniques.
When $V \ne 0$, the ground state now has two types of insulating phases.
The first one is the Mott phase where, similar to the on-site BH model, 
the ground-state boson occupancy is the same for every lattice site,
\textit{i.e.} $\langle \widehat{n}_i \rangle = n_0$. Here, 
$\langle ... \rangle$ is the thermal average, and the average boson occupancy
$n_0$ is chosen to minimize the ground-state energy for a given $\mu$.
The second one is the CDW phase which has crystalline order in 
the form of staggered boson occupancies, \textit{i.e.}
$\langle \widehat{n}_i \rangle = n_A$ and $\langle \widehat{n}_j \rangle = n_B$
for $i$ and $j$ nearest neighbors. 
To describe the CDW, it is convenient to 
split the entire lattice into two sublattices $A$ and $B$ such that the 
nearest-neighbor sites belong to a different sublattice. 
A lattice for which this can be done is called a bipartite lattice, 
and we assume the number of lattice sites in each sublattice is the 
same ($M/2$). We also assume that the boson occupancies of the 
sublattices $A$ and $B$ are $n_A$ and $n_B$, respectively, such that 
$n_A \ge n_B$. The case with $n_A = n_B = n_0$ corresponds to the Mott phase.

\begin{figure} [htb]
\centerline{\scalebox{0.5}{\includegraphics{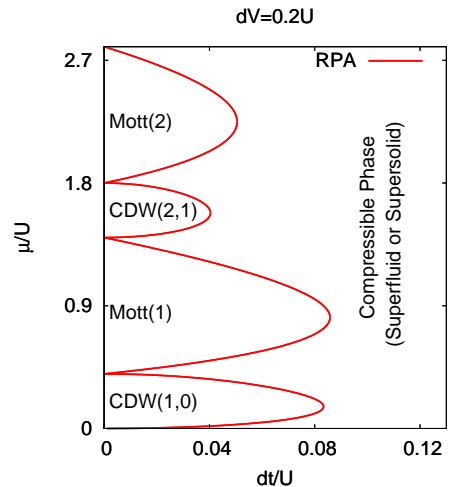}}}
\caption{\label{fig:mf} (Color online)
We show the chemical potential $\mu$ (in units of $U$) versus hopping $t$ 
(in units of $U/d$) phase diagram within the random-phase approximation
for $d$-dimensional hypercubic lattices (it becomes exact for $d \to \infty$).
Here the nearest-neighbor repulsion scales inversely with $d$ 
such that $dV = 0.2U$. The red solid line correspond to phase boundaries 
for the Mott insulator to superfluid and CDW insulator to supersolid 
states as obtained from Eq.~(\ref{eqn:mf-cdw}).
}
\end{figure}

When $t = 0$, it turns out that the chemical potential width of all 
Mott and CDW lobes are $U$ and $z V$, respectively, and that the ground 
state alternates between the CDW and Mott phases as a function of increasing 
$\mu$~\cite{bruder, parhat, otterlo, kuhner, kovrizhin, menotti, iskin}. 
For instance, the ground state is a vacuum 
$(n_A=0, n_B=0)$ for $\mu \le 0$; it is a CDW with
$(n_A=1, n_B=0) $ for $0 \le \mu \le z V$; it is a Mott insulator with $(n_A=1, n_B=1)$ 
for $zV \le \mu \le U+zV$; it is a CDW with $(n_A=2, n_B=1)$ 
for $U+zV \le \mu \le U + 2zV$; it is a Mott insulator with $(n_A=2, n_B=2)$ for 
$U+2zV \le \mu \le 2U+2zV$, and so on.
As $t$ increases, the range of $\mu$ about which the 
ground state is insulating decreases, and the Mott 
and CDW phases disappear at a critical value of $t$, beyond which 
the system becomes compressible (superfluid or supersolid) as shown in Fig.~\ref{fig:mf}.

Identification of these phases in atomic systems loaded into
optical lattices is a real challenge, and the momentum distribution 
of particles has been the most commonly used probing technique 
to distinguish superfluid and Mott phases of the on-site BH model. 
Motivated by these experiments, next we analyze the momentum 
distribution of the insulating phases of the extended BH model, paying particular 
attention to what signatures one might see that can distinguish the CDW insulating phase.

\section{Momentum Distribution}
\label{sec:md}

The momentum distribution of the atoms is one of the few (and probably 
the easiest) physical quantity that can be directly probed in 
experiments with ultracold atomic gases. This is achieved by 
time-of-flight absorption imaging of freely expanding atoms 
that are released from the trap. 
Since ultracold gases are very dilute, atoms do not interact
much with each other during this short time-of-flight, 
and therefore, the particle positions in the absorption image 
are strongly correlated with their velocity distribution given by
their momentum distribution at the moment of release from the trap.

The momentum distribution $n(\mathbf{k})$ is also easy to calculate, and it 
is defined as the Fourier transform of the one-particle density matrix
$
\rho(\mathbf{r},\mathbf{r'}) = \langle \psi^\dagger(\mathbf{r}) \psi(\mathbf{r'})  \rangle,
$
such that
\begin{equation}
\label{eqn:nk-gen}
n(\mathbf{k}) = \int d\mathbf{r} \int d\mathbf{r'} \rho(\mathbf{r}, \mathbf{r'}) e^{i\mathbf{k} \cdot (\mathbf{r}-\mathbf{r'})},
\end{equation}
where  $\psi^\dagger(\mathbf{r})$  [$\psi(\mathbf{r})$] is the boson 
creation (annihilation) field operator, and $\mathbf{k}$ is the momentum. 
We expand the field operators in the basis set of Wannier functions 
such that
$
\psi(\mathbf{r}) = (1/\sqrt{M}) \sum_\ell W(\mathbf{r}-\mathbf{R_\ell}) b_\ell,
$
where $M$ is the number of lattice sites, and the Wannier function 
$W(\mathbf{r}-\mathbf{R_\ell})$ is localized at site $\ell$ with 
position $\mathbf{R_\ell}$. 
Here the summation index $\ell \in \{A, B\}$ includes the entire lattice.

In this paper, we use two methods
to calculate the momentum distribution of the insulating phases of the 
extended BH model. First we calculate $n(\mathbf{k})$ via the RPA theory 
in Sec.~\ref{sec:rpa}, and its result corresponds to the exact result 
for the infinite-dimensional limit. Then,  in Sec.~\ref{sec:sc}, 
we calculate $n(\mathbf{k})$ as a power series expansion in the 
hopping $t$ via the strong-coupling perturbation theory. 
We also verify that our strong-coupling expansion recovers the RPA 
result in the infinite-dimensional limit when the latter is expanded out 
in $t$ to the same order. This provides an independent cross-check 
of the algebra as discussed next in detail.

\subsection{Random Phase Approximation (RPA)}
\label{sec:rpa}

Using the standard-basis operator method developed by 
Haley and Erd\"os~\cite{haley}, and following the recent works on
the on-site BH model~\cite{sheshadri, sengupta, menotti-rpa, konabe, ohashi},
here we obtain the equation of motion for the insulating phases of 
the extended BH model. This approximation is a well-defined linear operation 
in which thermal averages of products of operators are replaced by 
the product of their thermal averages. In accordance with this
approximation, the three-operator Green's functions are reduced
to two-operator ones~\cite{haley}. Therefore, the RPA method allows 
us to calculate the single-particle Green's function 
$
G(\mathbf{k}, i\omega_n)=-\langle \psi(\mathbf{k},i\omega_n) \psi^\dagger(\mathbf{k},i\omega_n) \rangle
$
in momentum ($\mathbf{k}$) and Matsubara frequency ($i\omega_n$) space,
from which the spectral function 
$
A(\mathbf{k}, \omega)=-(1/\pi) {\rm Im} G(\mathbf{k}, i\omega_n \to \omega+i\epsilon)
$
can be extracted by analytical continuation. Here the angular brackets denote the standard trace 
over the density matrix. 
Notice that the spectral function should always satisfy the 
sum rule
$
\int_{-\infty}^{\infty} A(\mathbf{k}, \omega) d\omega = 1,
$
due to the bosonic commutation relations of the creation and annihilation operators in the Heisenberg picture at equal times.
Then the momentum distribution 
$
n(\mathbf{k}) = \langle \psi^\dagger(\mathbf{k}) \psi(\mathbf{k}) \rangle
$
(at zero temperature) can be easily obtained from the 
spectral function
$
n(\mathbf{k}) = -\int_{-\infty}^0 A(\mathbf{k}, \omega) d\omega,
$
\textit{i.e.}
\begin{equation}
\label{eqn:nkrpa-def}
n(\mathbf{k})= \frac{1}{\pi} \int_{-\infty}^0 {\rm Im} G(\mathbf{k}, i\omega_n \to \omega+i\epsilon) d\omega,
\end{equation}
which measures the spectral weight of the hole excitation spectrum.

Expanding the field operators given in Eq.~(\ref{eqn:nk-gen}) in the basis 
set of Wannier functions, the momentum distribution becomes
\begin{equation}
\label{eqn:nksc-def}
n(\mathbf{k}) = \frac{|W(\mathbf{k})|^2}{M} \sum_{\ell, \ell'} 
\langle b_\ell^\dagger b_{\ell'} \rangle e^{-i\mathbf{k} \cdot (\mathbf{R_\ell}-\mathbf{R_{\ell'}})},
\end{equation}
where $W(\mathbf{k}) = \int d\mathbf{r} W(\mathbf{r}) e^{i\mathbf{k} \cdot \mathbf{r}}$
is the Fourier transform of $W(\mathbf{r})$. 
Here the summation indices $\ell \in \{A, B\}$ and $\ell' \in \{A,B\}$ include 
the entire lattice. Since $W(\mathbf{k})$ is
a nonuniversal property of the lattice potential, and it has
nothing to do with the extended BH model on a discrete periodic lattice, we ignore this 
function in this paper by setting it to unity. 

But before beginning the discussion of our formal treatment of the theory, 
we want to comment further on the subtle features that arise for the momentum 
distribution $n(\mathbf{k})$ in an ordered phase, when the lattice periodicity is 
further broken by the spontaneous appearance of the CDW phase 
with a lower lattice periodicity.  This system becomes that of
a lattice with a basis, as the $A$ and $B$ sublattices now have 
a different occupancies of particles on them. When examining $n(\mathbf{k})$ 
on the lattice, we evaluate the one-particle density matrix at each lattice 
site 
$
\rho({\bf r},{\bf r}^\prime)\rightarrow \rho({\bf R}_i,{\bf R}_j)=\rho_{ij}
$ 
(see the appendix for a further discussion of how one goes from the 
continuum to the lattice and how Wannier functions enter into the calculation). 
The integral in Eq.~(\ref{eqn:nk-gen}) is replaced by a summation 
{\it that extends over all lattice sites of the original lattice 
(before the CDW order occurred)}.  We can break this summation up 
into terms that involve solely the $A$ sublattice, 
solely the $B$ sublattice, and terms that mix the $A$ and $B$ sublattices.
One can immediately see that the terms restricted to one of the 
sublattices are periodic with the periodicity of the reduced 
Brillouin zone, while the mixed terms are only periodic
with respect to the full Brillouin zone. If we assume the Wannier 
functions are identical for the $A$ and $B$ sublattices, 
then this uniform weighting of the different contributions yields 
the correct momentum distribution; in general, one potentially 
has different weightings of the three different components.  
A full discussion of this issue is beyond this work, where we 
focus on the properties of the pure discrete lattice system, 
not on the experimental systems which have the additional real-space 
structure arising from the spatial continuum.

The fluctuations are not fully taken into account in the RPA method,
however it goes beyond the mean-field approximation for 
low-dimensional systems, and it becomes exact for 
infinite-dimensional bosonic systems recovering the mean-field theory. 
The RPA method has recently been applied to describe 
the superfluid and Mott phases of the on-site BH 
model~\cite{sengupta, menotti-rpa}, and its results showed good 
agreement with the experiments. Motivated by these earlier works, 
here we generalize this method to describe the insulating phases of 
the extended BH model.

Keeping in mind our two-sublattice system, the single-particle 
Green's function in momentum and frequency space can be written as
$
G(\mathbf{k}, i\omega_n) = (1/2) \sum_{S,S'} G_{SS'}(\mathbf{k}, i\omega_n),
$
where the indices $S$ and $S'$ label sublattices $\{A, B\}$, and
$
G_{SS'}(\mathbf{k}, i\omega_n) = (2/M) \sum_{\ell \in S, \ell' \in S'} 
G_{\ell\ell'}(i\omega_n) e^{-i\mathbf{k} \cdot (\mathbf{R_\ell}-\mathbf{R_{\ell'}})}
$
is the Fourier transform.  
Here the summation indices $\ell \in$ $A$ or $B$ and $\ell' \in$ $A$ or $B$ 
include only one sublattice and the Green's function is defined only at the different lattice positions. Since there are $M/2$ lattice sites in one sublattice, 
a factor of 2 appears in this expression.
Note that this is just a rewriting of the summation over all lattice sites that 
explicitly shows the contributions from the different sublattices.
The RPA equations have the following form in position and frequency space
$
G_{\ell\ell'}(i\omega_n) = G_\ell^0 (i\omega_n) \left[ \delta_{\ell\ell'} + \sum_{\ell''} J_{\ell\ell''} G_{\ell''\ell'}(i\omega_n) \right],
$
where $G_\ell^0 (i\omega_n)$ and $J_{\ell\ell''}$ are given 
below Eq.~(\ref{eqn:rpa-ins}). 
Here the indices $\ell$, $\ell'$ and $\ell''$ $\in \{A, B\}$ include the entire lattice.
Using the Fourier transforms, the RPA 
equation in momentum and frequency space becomes
$
G_{SS'}(\mathbf{k}, i\omega_n) = G_S^0 (\mathbf{k}, i\omega_n) \left[ \delta_{SS'} + \sum_{S''} J_{SS''}(\mathbf{k}) G_{S''S'}(\mathbf{k}, i\omega_n) \right].
$
This expression defines a set of coupled equations for the functions $G_{AA}(\mathbf{k}, i\omega_n)$,  
$G_{AB}(\mathbf{k}, i\omega_n)$,  $G_{BA}(\mathbf{k}, i\omega_n)$ and $G_{BB}(\mathbf{k}, i\omega_n)$.
These equations can be easily solved to obtain
\begin{widetext}
\begin{equation}
\label{eqn:rpa-gen}
G(\mathbf{k}, i\omega_n) = \frac{[G_A^0 (\mathbf{k}, i\omega_n) + G_B^0 (\mathbf{k}, i\omega_n) ]/2
+ [J_{AB}(\mathbf{k})+J_{BA}(\mathbf{k})-J_{AA}(\mathbf{k}) - J_{BB}(\mathbf{k})] 
G_A^0 (\mathbf{k}, i\omega_n) G_B^0 (\mathbf{k}, i\omega_n)/2}
{1 - J_{AA}(\mathbf{k}) G_A^0 (\mathbf{k}, i\omega_n) - J_{BB}(\mathbf{k}) G_B^0 (\mathbf{k}, i\omega_n)
-[J_{AB}(\mathbf{k}) J_{BA}(\mathbf{k})-J_{AA}(\mathbf{k}) J_{BB}(\mathbf{k})] 
G_A^0 (\mathbf{k}, i\omega_n) G_B^0 (\mathbf{k}, i\omega_n)},
\end{equation}
which corresponds to the general single-particle Green's function within the RPA.
\end{widetext}

We use Eq.~(\ref{eqn:rpa-gen}) to obtain the single-particle Green's function
for the insulating (Mott and CDW) phases of the extended BH model. 
Since hopping is allowed between nearest-neighbor sites that belong to different 
sublattices, in Eq.~(\ref{eqn:rpa-ins}) we find $J_{AA}(\mathbf{k}) = J_{BB}(\mathbf{k}) = 0$ 
and $J_{AB}(\mathbf{k}) = J_{BA}(\mathbf{k}) = \varepsilon(\mathbf{k})$, where
$\varepsilon(\mathbf{k})$ is the Fourier transform of the hopping matrix (also called the band structure).
For $d$-dimensional hypercubic lattices considered in this paper, the energy 
dispersion becomes
$
\varepsilon(\mathbf{k}) = -2t \sum_{i=1}^d \cos(k_i a),
$
where $a$ is the lattice spacing.  This then yields the following expression for the Green's function:
\begin{equation}
\label{eqn:rpa-ins}
G_{\rm Ins}(\mathbf{k}, i\omega_n) = \frac{G_{\rm avr}^0 (i\omega_n)
+ \varepsilon(\mathbf{k}) G_A^0 (i\omega_n) G_B^0 (i\omega_n)}
{1 - \varepsilon^2(\mathbf{k}) G_A^0 ( i\omega_n) G_B^0 (i\omega_n)},
\end{equation}
where $G_{\rm avr}^0 (i\omega_n) = \left[G_A^0 (i\omega_n) + G_B^0 (i\omega_n)\right]/2$.
The $\mathbf{k}$-independent functions $G_A^0 (i\omega_n)$ and 
$G_B^0 (i\omega_n)$ correspond to the single-particle local Green's functions 
for sublattices $A$ and $B$, respectively, at zeroth order in $t$. 
They have the familiar form
\begin{align}
G_A^0 (i\omega_n) &= \frac{n_A+1}{i\omega_n-E_A^{\rm par}} - \frac{n_A}{i\omega_n+E_A^{\rm hol}}, \\
\label{eqn:gb0}
G_B^0 (i\omega_n) &= \frac{n_B+1}{i\omega_n-E_B^{\rm par}} - \frac{n_B}{i\omega_n+E_B^{\rm hol}},
\end{align}
where 
$
E_A^{\rm par} = U n_A + z V n_B - \mu
$
and
$
E_B^{\rm par} = U n_B + z V n_A - \mu
$
are the zeroth-order particle excitation spectrum in $t$ 
(the energy required to add one extra particle) 
for sublattices $A$ and $B$, respectively, and similarly
$
E_A^{\rm hol} = -U (n_A-1) - z V n_B + \mu
$
and
$
E_B^{\rm hol} = -U (n_B-1) - z V n_A + \mu
$
are the zeroth-order hole excitation spectrum in $t$ 
(the energy required to remove one particle).
Notice that $G_{\rm Ins}(\mathbf{k}, i\omega_n) = G_{\rm avr}^0 (i\omega_n)$ at zeroth
order in $t$, as one may expect.

The poles of $G_{\rm Ins}(\mathbf{k}, i\omega_n)$, \textit{i.e.} the condition 
$1 = \varepsilon^2(\mathbf{k}) G_A^0 (i\omega_n) G_B^0 (i\omega_n)$, give the 
$\mathbf{k}$-dependence of the particle and hole excitation spectrum. 
The insulating phase becomes unstable against superfluidity when any of the 
excitation energies becomes negative at $\mathbf{k} = 0$. In addition, the poles of 
$G_{\rm Ins}(\mathbf{k}, i\omega_n)$ at $(\mathbf{k} = \mathbf{0}, i\omega_n = 0)$, 
\textit{i.e.} the condition $1 = \varepsilon^2(\mathbf{0}) G_A^0 (0) G_B^0 (0)$,
gives the mean-field phase boundary between the incompressible 
(Mott or CDW) and the compressible (superfluid or supersolid) phases.
This condition leads to
\begin{align}
\label{eqn:mf-cdw}
\frac{1}{z^2 t^2} &= \left( \frac{n_A+1}{E_A^{\rm par}} + \frac{n_A}{E_A^{\rm hol}} \right)
\left( \frac{n_B+1}{E_B^{\rm par}} + \frac{n_B}{E_B^{\rm hol}} \right),
\end{align}
which is a quartic equation for $\mu$, and it coincides with our earlier 
result~\cite{iskin}. Notice that Eq.~(\ref{eqn:mf-cdw}) reduces to the 
usual expression for the phase boundary of the on-site BH model 
when $n_A = n_B = n_0$ and $V = 0$. 
Having discussed the general RPA formalism for the insulating phases of 
the extended BH model, next we analyze the momentum distribution of 
the Mott and CDW phases separately.

\subsubsection{Mott Phase}
\label{sec:mi}

The single-particle Green's function for the Mott phase can be obtained 
from Eq.~(\ref{eqn:rpa-ins}) by setting $n_A = n_B = n_0$. This leads to
\begin{equation}
G_{\rm Mott}(\mathbf{k}, i\omega_n) = \frac{G_0^0 (i\omega_n)}
{1 - \varepsilon(\mathbf{k}) G_0^0 (i\omega_n)},
\end{equation}
which has the same form with that of the Green's function of the 
Mott phase in the on-site BH model~\cite{sengupta, ohashi, menotti-rpa}.
Here, $G_A^0 (i\omega_n) = G_B^0 (i\omega_n) = G_0^0 (i\omega_n)$.
The function $G_{\rm Mott}(\mathbf{k}, i\omega_n)$ has two poles at 
$i\omega_n = E_0^{\rm par}(\mathbf{k})$ and $i\omega_n = -E_0^{\rm hol}(\mathbf{k})$,
\begin{align}
E_0^{\rm par}(\mathbf{k}) &= E_0^{\rm par}-\left[U-\varepsilon(\mathbf{k})-E_0(\mathbf{k})\right]/2, \\
E_0^{\rm hol}(\mathbf{k}) &= E_0^{\rm hol}-\left[U+\varepsilon(\mathbf{k})-E_0(\mathbf{k})\right]/2,
\end{align}
corresponding to the particle (the energy required to add one extra particle) and hole 
(the energy required to remove one particle) excitation spectrum, respectively, where
$
E_0(\mathbf{k}) = \sqrt{\varepsilon^2(\mathbf{k})+2U(2n_0+1)\varepsilon(\mathbf{k})+U^2}.
$
Notice that the Mott insulator becomes unstable against superfluidity when 
$E_0^{\rm par}(\mathbf{0}) = 0$ or $E_0^{\rm hol}(\mathbf{0}) = 0$,
and these conditions coincide with the mean-field condition given in 
Eq.~(\ref{eqn:mf-cdw}) when $n_A = n_B = n_0$.

Therefore, the Green's function for the Mott phase can be written as
\begin{align}
\label{eqn:green-mott}
G_{\rm Mott}(\mathbf{k},i\omega_n) = \frac{C_0^{\rm par}(\mathbf{k})}{i\omega_n-E_0^{\rm par}(\mathbf{k})}
+ \frac{C_0^{\rm hol}(\mathbf{k})}{i\omega_n+E_0^{\rm hol}(\mathbf{k})},
\end{align}
where the coefficients (or the spectral weights) are functions of the excitation spectrum
\begin{align}
C_0^{\rm par}(\mathbf{k}) &= \frac{E_0^{\rm par}(\mathbf{k})+U n_0+E_0^{\rm hol}}
{E_0^{\rm par}(\mathbf{k})+E_0^{\rm hol}(\mathbf{k})}, \\
C_0^{\rm hol}(\mathbf{k}) &= \frac{E_0^{\rm hol}(\mathbf{k})-U n_0-E_0^{\rm hol}}
{E_0^{\rm par}(\mathbf{k})+E_0^{\rm hol}(\mathbf{k})}.
\end{align}
Using the definition given above Eq.~(\ref{eqn:nkrpa-def}), 
the spectral function for the Mott phase can be easily obtained 
from Eq.~(\ref{eqn:green-mott}), leading to
$
A_{\rm Mott}(\mathbf{k}, \omega) = C_0^{\rm par}(\mathbf{k}) \delta[\omega-E_0^{\rm par}(\mathbf{k})] 
+  C_0^{\rm hol}(\mathbf{k}) \delta[\omega+E_0^{\rm hol}(\mathbf{k})],
$
where $\delta(x)$ is the Delta function defined by
$
\delta(x) = (1/\pi) \lim_{\epsilon \to 0} \epsilon/(x^2 + \epsilon^2).
$
Notice that this function satisfies the sum rule mentioned above 
Eq.~(\ref{eqn:nkrpa-def}), since the coefficients
satisfy $C_0^{\rm par}(\mathbf{k})  + C_0^{\rm par}(\mathbf{k}) = 1$.
The momentum distribution measures the spectral weight of the hole excitation
spectrum as defined in Eq.~(\ref{eqn:nkrpa-def}), and for the Mott phase it is given by
\begin{equation}
\label{eqn:nkrpa-mott}
n_{\rm Mott}(\mathbf{k}) = - C_0^{\rm hol}(\mathbf{k}) 
= \frac{U(2n_0+1)+\varepsilon(\mathbf{k})}{2E_0(\mathbf{k})}-\frac{1}{2},
\end{equation}
which is identical to the $n_{\rm Mott}(\mathbf{k})$ of the on-site 
BH model~\cite{sengupta, menotti-rpa}.
Therefore, at the RPA level, $n_{\rm Mott}(\mathbf{k})$ is independent of 
$V$ which is mainly because of the underlying mean-field Hamiltonian 
that is used in the RPA formalism (we remind that fluctuations are 
not fully taken into account within RPA).
For instance, the mean-field phase boundary condition given in 
Eq.~(\ref{eqn:mf-cdw}) shows that the Mott lobes are separated by $z V$, 
but their shapes and, in particular, the critical points are independent of $V$.
This point will become more clear in Sec.~\ref{sec:sc}, where we analyze
$n(\mathbf{k})$ via the strong-coupling perturbation theory up to second order in $t$.
Notice that the momentum distribution is flat and equals the average filling fraction 
$n_{\rm Mott}(\mathbf{k}) = n_0$ at zeroth order in $t$, corresponding 
to vanishing site-to-site correlations.

\subsubsection{CDW Phase}
\label{sec:cdw}

In contrast to the Green's function of the Mott phase, the single-particle 
Green's function for the CDW phase $G_{\rm CDW}(\mathbf{k},i\omega_n)$ 
has four poles. Two of them correspond to the particle and the other 
two to the hole excitation spectrum of sublattices $A$ and $B$.
Unfortunately, general expressions for these poles are not 
analytically tractable since the condition
$
1 = \varepsilon^2(\mathbf{k}) G_A^0 (i\omega_n) G_B^0 (i\omega_n)
$ 
defines a quartic equation for $i\omega_n$; they can be easily 
obtained numerically for any given CDW lobe as shown in Sec.~\ref{sec:numerics}.
Assuming that the excitation spectrum is known, the Green's 
function for the CDW phase can be written as
\begin{align}
\label{eqn:green-cdw}
G_{\rm CDW}&(\mathbf{k},i\omega_n) = \frac{C_A^{\rm par}(\mathbf{k})}{i\omega_n-E_A^{\rm par}(\mathbf{k})}
+ \frac{C_A^{\rm hol}(\mathbf{k})}{i\omega_n+E_A^{\rm hol}(\mathbf{k})} \nonumber \\
&+ \frac{C_B^{\rm par}(\mathbf{k})}{i\omega_n-E_B^{\rm par}(\mathbf{k})}
+ \frac{C_B^{\rm hol}(\mathbf{k})}{i\omega_n+E_B^{\rm hol}(\mathbf{k})},
\end{align}
where $E_A^{\rm par}(\mathbf{k})$ and $E_B^{\rm par}(\mathbf{k})$ are the 
particle (the energy required to add one extra particle) and 
$E_A^{\rm hol}(\mathbf{k})$ and $E_B^{\rm hol}(\mathbf{k})$ are
the hole (the energy required to remove one particle) excitation spectrum.  
The coefficients (or the spectral weights) are functions of the excitation spectrum, such that
\begin{widetext}
\begin{align}
C_A^{\rm par}(\mathbf{k})= \frac{D_0(\mathbf{k})+D_1(\mathbf{k}) E_A^{\rm par}(\mathbf{k}) + D_2(\mathbf{k}) [E_A^{\rm par}(\mathbf{k})]^2 +  [E_A^{\rm par}(\mathbf{k})]^3}
{[E_A^{\rm par}(\mathbf{k})-E_B^{\rm par}(\mathbf{k})] [E_A^{\rm par}(\mathbf{k})+E_B^{\rm hol}(\mathbf{k})] [E_A^{\rm par}(\mathbf{k})+E_A^{\rm hol}(\mathbf{k})]}, \\
C_B^{\rm par}(\mathbf{k})= \frac{D_0(\mathbf{k})+D_1(\mathbf{k}) E_B^{\rm par}(\mathbf{k}) + D_2(\mathbf{k}) [E_B^{\rm par}(\mathbf{k})]^2 +  [E_B^{\rm par}(\mathbf{k})]^3}
{[E_B^{\rm par}(\mathbf{k})-E_A^{\rm par}(\mathbf{k})] [E_B^{\rm par}(\mathbf{k})+E_A^{\rm hol}(\mathbf{k})] [E_B^{\rm par}(\mathbf{k})+E_B^{\rm hol}(\mathbf{k})]}, \\
C_A^{\rm hol}(\mathbf{k})= \frac{D_0(\mathbf{k})-D_1(\mathbf{k}) E_A^{\rm hol}(\mathbf{k}) + D_2(\mathbf{k}) [E_A^{\rm hol}(\mathbf{k})]^2 -  [E_A^{\rm hol}(\mathbf{k})]^3}
{[E_B^{\rm hol}(\mathbf{k})-E_A^{\rm hol}(\mathbf{k})] [E_A^{\rm hol}(\mathbf{k})+E_B^{\rm par}(\mathbf{k})] [E_A^{\rm hol}(\mathbf{k})+E_A^{\rm par}(\mathbf{k})]}, \\
C_B^{\rm hol}(\mathbf{k})= \frac{D_0(\mathbf{k})-D_1(\mathbf{k}) E_B^{\rm hol}(\mathbf{k}) + D_2(\mathbf{k}) [E_B^{\rm hol}(\mathbf{k})]^2 -  [E_B^{\rm hol}(\mathbf{k})]^3}
{[E_A^{\rm hol}(\mathbf{k})-E_B^{\rm hol}(\mathbf{k})] [E_B^{\rm hol}(\mathbf{k})+E_B^{\rm par}(\mathbf{k})] [E_B^{\rm hol}(\mathbf{k})+E_A^{\rm par}(\mathbf{k})]}.
\end{align}
\end{widetext}
Here, the coefficients $D_0(\mathbf{k})$, $D_1(\mathbf{k})$, and $D_2(\mathbf{k})$ are 
functions of the zeroth-order excitation spectrum in $t$ defined below Eq.~(\ref{eqn:gb0}), and are given by
\begin{align}
D_0&(\mathbf{k}) = -\left[ E_A^{\rm par} E_A^{\rm hol} (U n_B+E_B^{\rm hol}) + E_B^{\rm par} E_B^{\rm hol} \right. \nonumber\\
&\left. (U n_A+E_A^{\rm hol}) \right]/2 + \varepsilon(\mathbf{k}) (U n_A+E_A^{\rm hol}) (U n_B+E_B^{\rm hol}),
\end{align}
\begin{align}
D_1&(\mathbf{k}) = \left[ (E_A^{\rm hol}-E_A^{\rm par}) (U n_B+E_B^{\rm hol}) + (E_B^{\rm hol}-E_B^{\rm par}) \right.\nonumber\\
&\left. (U n_A+E_A^{\rm hol}) - E_A^{\rm par} E_A^{\rm hol} - E_B^{\rm par} E_B^{\rm hol} \right]/2 \nonumber \\
&+ \varepsilon(\mathbf{k}) (U n_A+U n_B+E_A^{\rm hol}+E_B^{\rm hol}),
\end{align}
and
\begin{align}
D_2&(\mathbf{k}) = \left(U n_A+U n_B-E_A^{\rm par}-E_B^{\rm par}\right)/2+E_A^{\rm hol} +E_B^{\rm hol} \nonumber\\
& + \varepsilon(\mathbf{k}).
\end{align}

Using the definition given above Eq.~(\ref{eqn:nkrpa-def}), the spectral function 
for the CDW phase can be easily obtained from Eq.~(\ref{eqn:green-cdw}), leading to
$
A_{\rm CDW}(\mathbf{k}, \omega) 
= C_A^{\rm par}(\mathbf{k}) \delta[\omega-E_A^{\rm par}(\mathbf{k})] 
+ C_A^{\rm hol}(\mathbf{k}) \delta[\omega+E_A^{\rm hol}(\mathbf{k})]
+ C_B^{\rm par}(\mathbf{k}) \delta[\omega-E_B^{\rm par}(\mathbf{k})] 
+ C_B^{\rm hol}(\mathbf{k}) \delta[\omega+E_B^{\rm hol}(\mathbf{k})].
$
Notice that this function satisfies the sum rule mentioned above 
Eq.~(\ref{eqn:nkrpa-def}), since the coefficients satisfy
$
C_A^{\rm par}(\mathbf{k}) + C_A^{\rm hol}(\mathbf{k}) 
+ C_B^{\rm par}(\mathbf{k}) + C_B^{\rm hol}(\mathbf{k}) =1.
$
The momentum distribution measures the spectral weight of the hole excitation
spectrum as defined in Eq.~(\ref{eqn:nkrpa-def}), and for the CDW phase it is given by
\begin{equation}
\label{eqn:nkrpa-cdw}
n_{\rm CDW}(\mathbf{k}) = - C_A^{\rm hol}(\mathbf{k}) -  C_B^{\rm hol}(\mathbf{k}).
\end{equation}
This expression has a highly non-trivial dependence on $t$, and it has to be
solved numerically together with the excitation spectrum. However, 
it can be analytically shown that the momentum distribution is 
flat and equals the average filling fraction 
$n_{\rm CDW}(\mathbf{k}) = (n_A+n_B)/2$ at zeroth order in $t$, corresponding 
to vanishing site-to-site correlations. To provide an independent check of 
the algebra (and to extend to finite dimensions), we next calculate $n(\mathbf{k})$ 
as a power series expansion in the hopping $t$ via the exact strong-coupling perturbation theory in $d$ dimensions.

\subsection{Strong-coupling Perturbation Theory}
\label{sec:sc}

To determine the momentum distribution of the insulating phases, 
we need the wavefunction of the insulating state $| \Psi_{\rm Ins} \rangle$ 
as a function of $t$. We use the many-body version of 
Rayleigh-Schr\"odinger perturbation theory in the kinetic 
energy term~\cite{landau} to perform the expansion 
(in powers of $t$) for $| \Psi_{\rm Ins} \rangle$ needed to 
carry out our analysis. A similar expansion for the ground-state energies 
was previously used to discuss the phase diagram of the on-site BH 
model~\cite{freericks-1, freericks-2}, and it has recently been applied 
to the extended BH model~\cite{iskin}. For the on-site BH model, extrapolated 
results of these expansions showed an excellent agreement with 
recent quantum Monte Carlo simulations~\cite{prokofiev-1,prokofiev-2}.
A high-order strong-coupling expansion for the ground-state energies 
has now been extended to all dimensions and fillings~\cite{holthaus}, 
and a high-order expansion for the wavefunction has also been used 
to describe the Mott phase in one-dimensional systems~\cite{damski}. 

For our purpose, we first need the ground-state wavefuntions of the 
Mott and CDW phases when $t = 0$. To zeroth order in $t$, the insulator 
(Mott or CDW) wavefunction can be written as
\begin{eqnarray}
\label{eqn:wf-0}
|\Psi_{\rm Ins}^{(0)} \rangle &=& \prod_{\ell \in A, \ell' \in B}^{M/2} 
\frac{(b_\ell^\dagger)^{n_A}}{\sqrt{n_A!}} \frac{(b_{\ell'}^\dagger)^{n_B}}{\sqrt{n_B!}} | 0 \rangle,
\end{eqnarray}
where $M$ is the number of lattice sites, and $| 0 \rangle$ is the vacuum state 
(here, we remind that the lattice is divided equally into $A$ and $B$ sublattices). 
In principle, we can apply the perturbation theory on $|\Psi_{\rm Ins}^{(0)} \rangle$ 
to calculate $|\Psi_{\rm Ins} \rangle$ up to the desired order. 
However, since the number of intermediate states increases 
dramatically due to the presence of nearest-neighbor interactions, 
we perform this expansion only up to second order in $t$.
The (unnormalized) wavefunction for the insulating
state can then be written as
\begin{align}
\label{eqn:wf-non}
|\psi_{\rm Ins} \rangle &= | \Psi_{\rm Ins}^{(0)} \rangle
+ \sum_{m \ne | \Psi_{\rm Ins}^{(0)} \rangle} \frac{T_{m0}}{E_{0m}} | \Psi_{\rm Ins}^{(0)} \rangle \nonumber \\
&+  \sum_{\{m',m\} \ne | \Psi_{\rm Ins}^{(0)} \rangle} \frac{T_{m'm} T_{m0}}{E_{0m'} E_{0m}} | \Psi_{\rm Ins}^{(0)} \rangle + O(t^3),
\end{align}
where 
$
T_{m0} = -\sum_{S,S'} \sum_{\ell \in S, \ell' \in S'} t_{\ell \ell'} \langle m | b_\ell^\dagger b_{\ell'} | \Psi_{\rm Ins}^{(0)} \rangle
$ 
is the hopping matrix element between the first-order intermediate state $| m \rangle$ 
and the zeroth-order state $| \Psi_{\rm Ins}^{(0)} \rangle$, and $T_{mm'}$ 
is between $| m \rangle$ and the second-order intermediate state $| m' \rangle$, 
and $E_{0m} = E_{\rm Ins}^{(0)}-E_m^{(0)}$. 
Here the summation indices $\ell \in \{A,B\}$ and $\ell' \in \{A,B\}$ 
include the entire lattice, and $S$ and $S'$ label sublattices $\{A, B\}$. 
The $| m \rangle$ states 
are connected to $| \Psi_{\rm Ins}^{(0)} \rangle$ state with a single hopping, 
and similarly $| m' \rangle$ states are connected to $| m \rangle$ states with a single 
hopping. However, the $| m' \rangle$ state must be different from the $| \Psi_{\rm Ins}^{(0)} \rangle$ state.

To calculate the momentum distribution, we need the normalized wavefunction 
for the insulating state
$
|\Psi_{\rm Ins} \rangle = |\psi_{\rm Ins} \rangle  / \sqrt{\langle \psi_{\rm Ins} |\psi_{\rm Ins} \rangle},
$
where the normalization up to second order in $t$ is given by
\begin{align}
\langle &\psi_{\rm Ins} | \psi_{\rm Ins} \rangle = 1 
+ \frac{n_A (n_B+1) M z t^2/2}{\left[U(n_A-n_B-1)+V(z n_B-z n_A+1)\right]^2} \nonumber \\
&+ \frac{n_B (n_A+1) M z t^2/2}{\left[U(n_B-n_A-1)+V(z n_A-z n_B+1)\right]^2} + O(t^4).
\end{align}
Here, $z = 2d$ is the lattice coordination number.
Since $\langle m | \Psi_{\rm Ins}^{(0)} \rangle = \langle m'' | \Psi_{\rm Ins}^{(0)} \rangle = \langle m'|m \rangle = 0$,
the first and third order terms in $t$ vanish in the normalization. 
In general, all odd-order terms in $t$ vanish.

A lengthy but straightforward calculation leads to the momentum distribution, 
[defined in Eq.~(\ref{eqn:nksc-def})]
$
n(\mathbf{k}) = (1/M) \sum_{\ell, \ell'} \langle \Psi_{\rm Ins} | b_\ell^\dagger b_{\ell'} | \Psi_{\rm Ins} \rangle  
e^{-i\mathbf{k} \cdot (\mathbf{R_\ell}-\mathbf{R_{\ell'}})},
$ 
up to second order in $t$ as
\begin{widetext}
\begin{align}
\label{eqn:nksc}
n_{\rm Ins}(\mathbf{k}) &= \frac{n_A+n_B}{2}
+ \left[ \frac{n_A (n_B+1)}{U(n_A-n_B-1)+V(z n_B-z n_A+1)} + \frac{n_B (n_A+1)}{U(n_B-n_A-1)+V(z n_A-z n_B+1)} \right] \varepsilon(\mathbf{k}) \nonumber \\
&+ \Big\lbrace \frac{n_A (n_B+1)}{2\left[U(n_A-n_B-1)+V(z n_B-z n_A+1)\right]^2} + \frac{n_B (n_A+1)}{2\left[U(n_B-n_A-1)+V(z n_A-z n_B+1)\right]^2} \nonumber \\
&- \frac{n_A (n_B+1)}{U\left[U(n_A-n_B-1)+V(z n_B-z n_A+1)\right]} - \frac{n_B (n_A+1)}{U\left[U(n_B-n_A-1)+V(z n_A-z n_B+1)\right]} \Big\rbrace \nonumber \\
& (n_A+n_B+1) \left[\varepsilon^2(\mathbf{k})-2dt^2\right] + O(t^3).
\end{align}
\end{widetext}
In the definition of the momentum distribution, the summation indices $\ell \in \{A, B\}$ 
and $\ell' \in \{A,B\}$ include the entire lattice.
Here
$
\varepsilon(\mathbf{k}) = - (2/M) \sum_{\ell \in S, \ell' \in S'} t_{\ell \ell'} e^{i\mathbf{k} \cdot (\mathbf{R_\ell}-\mathbf{R_{\ell'}})}
$
is the Fourier transform of the hopping matrix $t_{\ell,\ell'}$ (energy dispersion), and
$
\varepsilon^2(\mathbf{k})-2dt^2 = (2/M) \sum_{\{\ell,\ell''\} \in S, \ell' \in S'} t_{\ell \ell'} t_{\ell' \ell''} e^{i\mathbf{k} \cdot (\mathbf{R_\ell}-\mathbf{R_{\ell''}})},
$
where the summation indices $\{\ell, \ell''\} \in$ $A$ (or $B$) and $\ell' \in$ $B$ 
(or $A$) include only one sublattice.
Since there are $M/2$ lattice sites in one sublattice,  a factor of 2
appears in these expressions.
To zeroth order in $t$, Eq.~(\ref{eqn:nksc}) 
shows that $n_{\rm Ins}(\mathbf{k})$ is flat and equals the average filling 
fraction $(n_A+n_B)/2$. However, it develops a peak around 
$\mathbf{k} = \mathbf{0}$ and a minimum around $\mathbf{k} = \mathbf{\pi}$  
at first order in $t$. These general observations are consistent with the RPA results
shown in Eqs.~(\ref{eqn:nkrpa-mott}) and~(\ref{eqn:nkrpa-cdw}).

Equation~(\ref{eqn:nksc}) is valid for the insulating phases of all 
$d$-dimensional hypercubic lattices. For instance, when $n_A = n_B = n_0$, 
Eq.~(\ref{eqn:nksc}) reduces to the momentum distribution for
the Mott phase, i.e.
\begin{align}
\label{eqn:nksc-mott}
n_{\rm Mott}&(\mathbf{k}) = n_0 - 2n_0(n_0+1) \frac{\varepsilon(\mathbf{k})}{U-V} + n_0(n_0+1) \nonumber \\ 
& (2n_0+1) \left [\varepsilon^2(\mathbf{k})-2dt^2\right]\frac{3U-2V}{U(U-V)^2} + O(t^3).
\end{align}
This expression recovers the known result for the on-site BH model 
when $V = 0$~\cite{freericks-3, hoffmann}.
In addition, in the $d \to \infty$ limit, we checked that Eqs.~(\ref{eqn:nksc}) 
and~(\ref{eqn:nksc-mott}) agree with the RPA solutions (which are exact in this limit) 
given in Eqs.~(\ref{eqn:nkrpa-cdw}) and~(\ref{eqn:nkrpa-mott}) when 
the latter are expanded out to second order in $t$, providing an independent 
check of the algebra. One must note that the terms $2V$ and $V$
that appear in the numerator and denominator of Eq.~(\ref{eqn:nksc-mott}) vanish 
in the limit when $d \to \infty$ because $V \propto 1/d$.
Next we compare the RPA results with those of the strong-coupling
perturbation theory.

\begin{figure} [htb]
\centerline{\scalebox{0.5}{\includegraphics{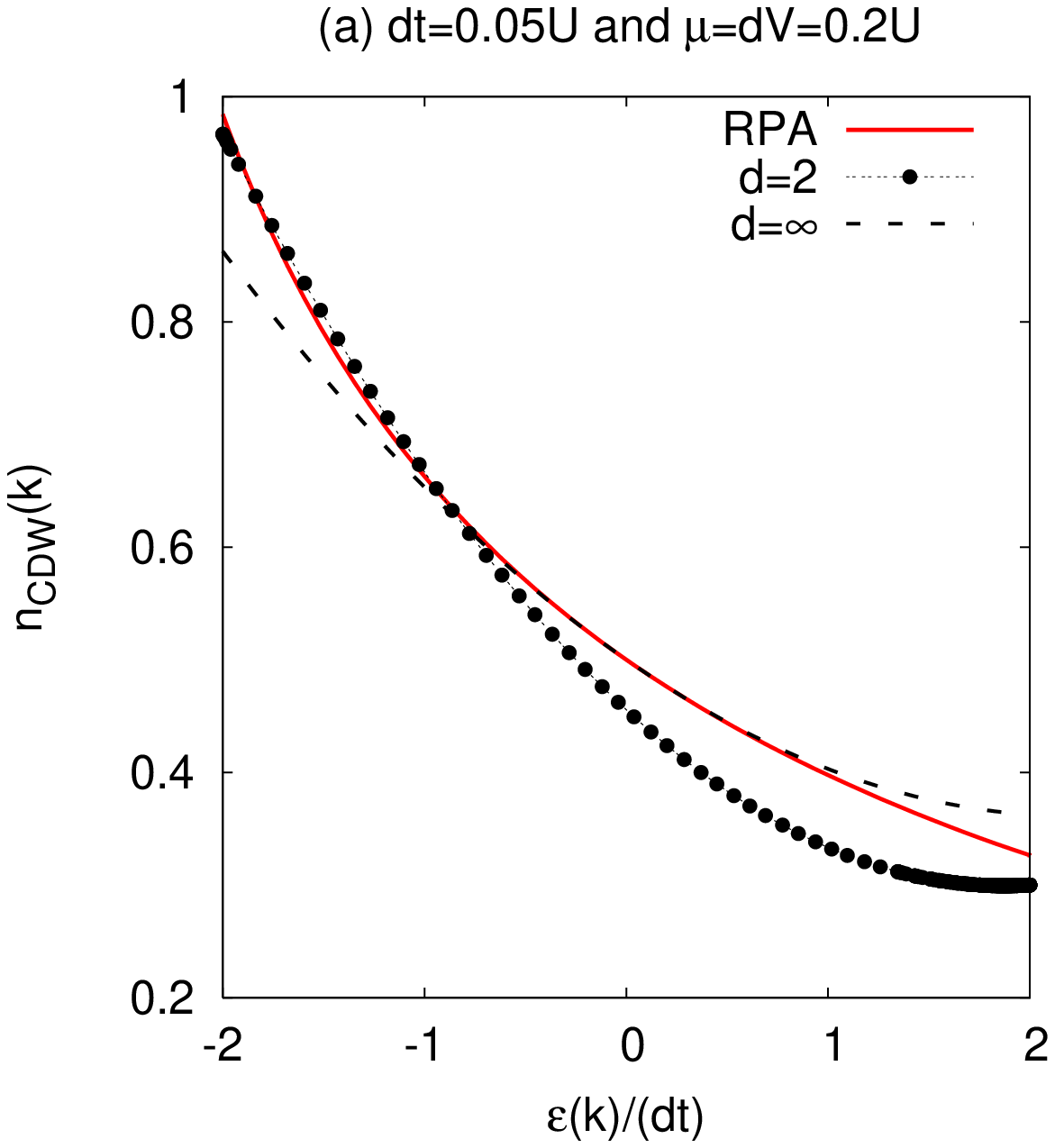}}}
\centerline{\scalebox{0.5}{\includegraphics{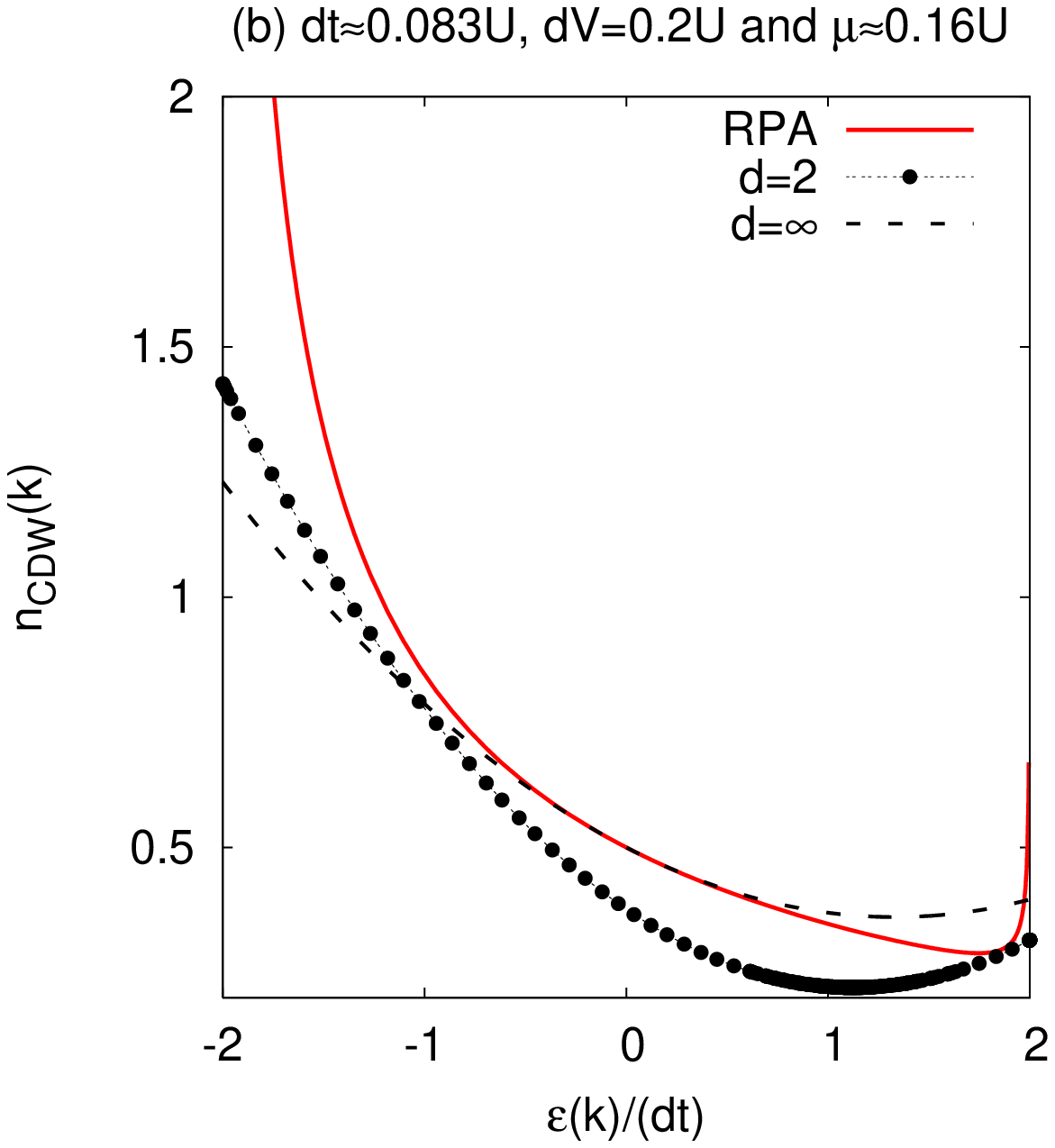}}}
\caption{\label{fig:Ek} (Color online)
Momentum distribution $n_{\rm CDW}(\mathbf{k})$ versus 
$\varepsilon(\mathbf{k})/(dt)$ for a $(d=2)$- and $(d \to \infty)$-dimensional 
hypercubic lattices.
Panel (a) has the nearest-neighbor boson-repulsion satisfying $dV = 0.2U$, 
the hopping satisfying $dt = 0.05U$ and the chemical potential set by $\mu = 0.2U$ 
corresponding approximately to the center of the first CDW lobe, and panel
(b) has $dV = 0.2U$, $dt \approx 0.083U$, and $\mu \approx 0.16$ 
corresponding approximately to the tip of the first CDW lobe.
The solid (red) lines correspond to the RPA, and the dashed and circled 
lines to the second-order strong-coupling perturbation theory 
for different dimensions. The peak occurs at the zone corner only when the hopping is close to the tip of the CDW lobe.
}
\end{figure}
\subsection{Numerical Results}
\label{sec:numerics}

Since the momentum distribution of the CDW phase given in 
Eq.~(\ref{eqn:nkrpa-cdw}) has a highly nontrivial dependence 
on $t$, it has to be solved numerically together with the excitation spectrum.
Next we set $dV = 0.2U$ and solve this equation for the first CDW lobe.
For this parameter, we remind that the $t=0$ chemical potential width of 
all Mott and CDW lobes are $U$ and $0.4U$, respectively, and that the ground 
state alternates between the CDW and Mott phases as a function of $\mu$. 
For instance, the ground state is a vacuum 
$(n_0=0) $ for $\mu \le 0$; it is a CDW with
$(n_A=1, n_B=0)$ for $0 \le \mu \le 0.4U$; it is a Mott insulator with $(n_0=1)$ 
for $0.4U \le \mu \le 1.4U$; it is a CDW with $(n_A=2, n_B=1)$ 
for $1.4U \le \mu \le 1.8U$; it is a Mott insulator with $(n_0=2)$ for 
$1.8U \le \mu \le 2.8U$.

In Fig.~\ref{fig:Ek}, the results of the RPA calculation given in 
Eq.~(\ref{eqn:nkrpa-cdw}) are compared to those of the second-order 
strong-coupling perturbation theory given in Eq.~(\ref{eqn:nksc}) 
for a $(d=2)$- and $(d \to \infty)$-dimensional hypercubic lattices.
In this figure, we show the momentum distribution $n_{\rm CDW}(\mathbf{k})$ 
as a function of $\varepsilon(\mathbf{k})/(dt)$ for two sets of parameters. 
In Fig.~\ref{fig:Ek}(a), we choose $dt = 0.05U$ 
and $\mu = 0.2U$ which approximately corresponds to the center of the first CDW lobe. 
For this parameter set, deep inside the CDW lobe, the momentum distribution 
has a peak at $\varepsilon(\mathbf{k})=-2dt$ corresponding to the $\mathbf{k} = \mathbf{0}$ point,
and it has a minimum at $\varepsilon(\mathbf{k})=2dt$ corresponding to the 
$\mathbf{k} = (\pi,\pi,\ldots)$ point. This is very similar to what happens in the Mott phase.
However, in Fig.~\ref{fig:Ek}(b), we choose 
$dt \approx 0.083U$ and $\mu \approx 0.16U$ which approximately 
corresponds to the tip of the first CDW lobe. 
For this parameter set, close to the CDW-supersolid phase transition, 
the momentum distribution has two peaks: a large peak at 
$\varepsilon(\mathbf{k})=-2dt$ corresponding to the $\mathbf{k} = \mathbf{0}$ 
point, and a smaller one at $\varepsilon(\mathbf{k})=2dt$ corresponding 
to the $\mathbf{k} = (\pi,\pi,\ldots)$ point. The second peak is unique to the 
CDW phase and it does not occur in a Mott phase. 
Notice that both the RPA and second-order strong-coupling 
expansion give qualitatively similar results (although the peak is much 
sharper and has lower weight in the exact solution).

One might have expected to always see the peak in the momentum distribution 
at the $\mathbf{k} = (\pi,\pi,\ldots)$ point due to the reduced periodicity of the CDW order. 
But because the momentum distribution involves four terms corresponding the the $AA$, 
$AB$, $BA$, and $BB$ sublattice combinations, only the first and last 
terms are periodic in the reduced Brillouin zone (see our discussion
given in the appendix). Deep inside the CDW lobe, the presence of 
a large gap in the one-particle excitation spectrum produces 
an exponential decay of the one-particle correlations
which suppresses this peak in the momentum distribution as can be 
seen in Fig.~\ref{fig:Ek}(a) (this point has already been discussed in  Ref.~\onlinecite{rigol}).
This essentially occurs because there is a cancellation of the peak that 
arises from the $AA$ and $BB$ contributions with the results from 
the $AB$ and $BA$ pieces, similar to what happens in the Mott phase.
However, close to the tip of the CDW lobe, the peak emerges 
in the exact solution of the RPA as shown in Fig.~\ref{fig:Ek}(b).
To some extent, this peak also emerges in the solutions of the
second-order strong-coupling perturbation theory. Notice that the
peak is underemphasized in the strong-coupling theory since the 
theory is exact only deep inside the CDW lobe, and it becomes 
quantitatively inaccurate for large values of $dt/U$ close 
to the tip of the CDW lobe. We remark that an unphysical peak
appears at $\mathbf{k} = (\pi,\pi,\ldots)$ in the strong-coupling 
perturbation theory for the Mott phase (not shown), which signals 
the breakdown of the second-order expansion.

\begin{figure} [htb]
\centerline{\scalebox{0.5}{\includegraphics{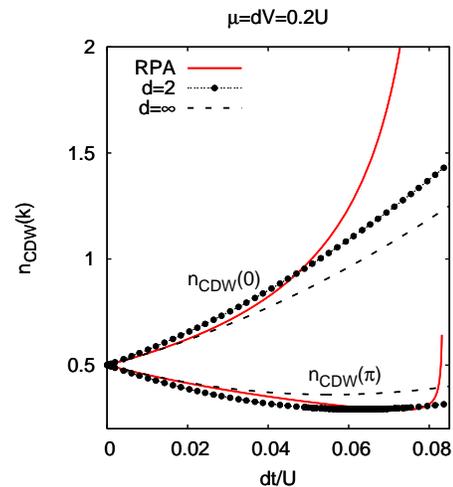}}}
\caption{\label{fig:t} (Color online)
Momentum ditributions at specific momentum points $n_{\rm CDW}(\mathbf{k} = \mathbf{0})$ and 
$n_{\rm CDW}(\mathbf{k} = \mathbf{\pi})$ versus $dt/U$ 
for $(d=2)$- and $(d \to \infty)$-dimensional hypercubic lattices.
The chemical potential $\mu = dV$ corresponds to the first CDW lobe, 
and the nearest-neighbor repulsion is set to $dV = 0.2U$.
The solid line corresponds to the RPA and the dashed and circled lines 
to the second-order strong-coupling perturbation theory 
for different dimensions.
}
\end{figure}

As a further check of the accuracy of our second-order strong-coupling 
expansion, in Fig.~\ref{fig:t} we compare the $d=2$ and $d \to \infty$ limits of 
Eq.~(\ref{eqn:nksc}) to the RPA method given in Eq.~(\ref{eqn:nkrpa-cdw}) 
which corresponds to the exact solution in the latter limit. 
In this figure, we show $n_{\rm CDW}(\mathbf{k} = \mathbf{0})$ and 
$n_{\rm CDW}(\mathbf{k} = \mathbf{\pi})$ as a function 
of $dt/U$ when $\mu = 0.2U$.
In $d=2$ dimensions, the RPA and second-order strong-coupling expansion 
gives qualitatively similar results for small values of $dt/U$, 
\textit{i.e.} deep inside the CDW lobe. 
However, in the $d \to \infty$ limit, the results of the RPA and the second-order 
strong-coupling expansion match exactly for small values of $dt/U$ (as they must).
Close to the tip of the CDW lobe, the RPA and strong-coupling 
results differ substantially from each other signalling the 
breakdown of the second-order expansion.
However, both theories show that $n_{\rm CDW}(\mathbf{0})$ 
is an increasing function of $dt/U$ as one may expect. 
This is because the range of $\mu$ about which the ground state is a CDW 
decreases as $dt/U$ increases from zero, and the CDW phase become
a supersolid at a critical value of $d t_c \sim 0.08U$. 
Beyond this point, $n(\mathbf{0})$ diverges due to the appearance of a condensate,
corresponding to the macroscopic occupation of the $\mathbf{k} = \mathbf{0}$ state.

Note that we do not attempt to perform a scaling analysis of the momentum 
distribution for the CDW phase. The reasons why are twofold. 
First, we only have the series through second order, which probably is too 
short to be able to properly fit to a phenomenological scaling form, and second, 
we cannot extract the analytic scaling form from the RPA calculation anymore, 
so guessing an appropriate phenomenological form has less guidance than for the
Mott phase. A scaled theory would be expected to be accurate for all values of $t$ 
within the insulating phases, as has been recently shown for 
the Mott phase of the on-site BH model~\cite{freericks-3}.

\section{Conclusions}
\label{sec:conclusions}

We developed two methods to calculate the momentum distribution of the
insulating (Mott and charge-density-wave) phases of the extended 
Bose-Hubbard model with on-site and nearest-neighbor boson-boson 
repulsions on $d$-dimensional hypercubic lattices. 
First we analyzed the momentum distribution within the random 
phase approximation, which corresponds to the exact 
solution for the infinite-dimensional limit.
Then we used the many-body version of the Rayleigh-Schr\"odinger 
perturbation theory in the kinetic-energy term, and derived the 
wavefunction for the insulating phases as a power series in the hopping $t$,
to calculate the momentum distribution via the strong-coupling perturbation 
theory. A similar strong-coupling expansion for the ground-state energies 
was previously used to discuss the phase diagram of the on-site BH 
model~\cite{freericks-1, freericks-2}, and it has recently been applied 
to the extended BH model~\cite{iskin}.

The agreement between the second-order strong-coupling expansion 
and that of RPA method is only qualitative in low-dimensional systems. This is 
not surprising since the fluctuations are not fully taken into account
in the RPA method. However, we showed that our strong-coupling 
expansion matches exactly the RPA result (as it must) 
in the infinite-dimensional limit when the latter is expanded out 
in $t$ to the same order. We believe some of these results could potentially 
be tested with ultracold dipolar Bose gases loaded into optical lattices. 
This work can be extended in several ways if desired. For instance,
one could calculate the momentum distribution up to third order in $t$,
and develop a scaling theory with the help of the RPA results 
(or a good phenomenological guess for the scaling form of the momentum distribution). 
The scaled theory is expected to be accurate for all values of $t$ 
within the insulating phases, as has been recently shown for 
the Mott phase of the on-site BH model~\cite{freericks-3}.

\section{Acknowledgements}
\label{sec:ack}
We would like to acknowledge useful discussions with H. R. Krishnamurthy and M. Rigol.
J. K. F. acknowledges support under the USARO Grant W911NF0710576 with
funds from the DARPA OLE Program.

\appendix
\section{Effective CDW Hamiltonian}
\label{sec:app}

In this Appendix, we comment on some of the subtle issues regarding Wannier 
functions in the CDW phase. When the particle occupancies show a CDW order, 
we can think of the combination of the CDW order plus the lattice potential 
as an effective lattice potential such that the effective potential is different 
for each sublattice. In other words, CDW order creates an effective 
potential which depends on the particle occupation of the sublattice. 
In fact, having different effective lattice potentials on two sublattices 
could be thought of as the reason for having a CDW order at the first place. Equivalently, this 
is like considering the mean-field Hamiltonian with CDW order as the starting point for
determining the Wannier wavefunctions, with the symmetry explicitly broken between the
$A$ and $B$ sublattices.

This observation suggests that in contrast to the Mott phase where all lattice
sites are identical and the Wannier functions are exactly the same for
both sublattices, \textit{i.e.} $W_A(\mathbf{r}) = W_B(\mathbf{r}) = W_0(\mathbf{r})$,  
the Wannier functions depend on the sublattice when the CDW order 
exists, \textit{i.e.} $W_A(\mathbf{r}) \ne W_B(\mathbf{r})$. 
Throughout this paper, we assume that the Wannier 
functions are equal (or at least similar) in sublattices $A$ and $B$. 
However, depending on the CDW order (\textit{e.g.} $n_A \gg n_B$) 
and the lattice potential, the Wannier functions of one sublattice may 
become substantially different from that of the other.  
In such a case, the field operator can be expanded as
$
\psi(\mathbf{r}) = (1/\sqrt{M}) \sum_S \sum_{\ell \in S} W_S(\mathbf{r}-\mathbf{R_\ell}) b_\ell,
$
where $M$ is the number of lattice sites, $S$ labels sublattices $\{A, B\}$, and 
$
W_S(\mathbf{k}) = \int d\mathbf{r} W_S(\mathbf{r}) e^{i\mathbf{k} \cdot \mathbf{r}}
$
is the Fourier transform. 
Here the summation index $\ell \in \{A, B\}$ includes the entire lattice.

When $W_A(\mathbf{r}) \ne W_B(\mathbf{r})$, the strength of the on-site 
boson-boson repulsion also depends on the sublattice, 
since the effective interaction
$
U_S^{\rm eff} = g \int d\mathbf{r} |W_S(\mathbf{r})|^4
$
is larger for deeper potentials, where $g$ is the bare boson-boson repulsion
of the continuum Hamiltonian. Therefore, the effective Hamiltonian that describes 
the CDW phase can be written as 
\begin{align}
\label{eqn:cdwh}
H_{\rm CDW}^{\rm eff} &= - t^{\rm eff} \sum_{\langle i,j \rangle} b_i^\dagger b_j 
- \mu \sum_{i \in \{A,B\}} \widehat{n}_i \nonumber \\ 
&+ \frac{U_A^{\rm eff}}{2} \sum_{i \in A} \widehat{n}_i (\widehat{n}_i-1)
+ \frac{U_B^{\rm eff}}{2} \sum_{j \in B} \widehat{n}_j (\widehat{n}_j-1) \nonumber \\
&+ V_{\rm AB}^{\rm eff} \sum_{\langle i \in A,j \in B \rangle} \widehat{n}_i \widehat{n}_j,
\end{align}
where the notation $\langle i, j \rangle$ corresponds to nearest-neighbors.
Here the effective hopping element 
$t_{AB}^{\rm eff} = t_{BA}^{\rm eff} = t^{\rm eff}$ between the 
two sublattices is given by
$
t^{\rm eff} = - \int d\mathbf{r} W_A^*(\mathbf{r}-\mathbf{R_i})
[ - \nabla^2 /(2m)  + V_{\rm OL}(\mathbf{r}) ] W_{B}(\mathbf{r}-\mathbf{R_j}),
$
where $m$ is the mass of particles, and $V_{\rm OL}(\mathbf{r})$ is the 
lattice potential, and
$
V_{AB}^{\rm eff} = g \int d\mathbf{r} |W_A(\mathbf{\mathbf{r}-\mathbf{R_i}})|^2|W_B(\mathbf{\mathbf{r}-\mathbf{R_j}})|^2
$
is the effective nearest-neighbor boson-boson repulsion. 

When $W_A(\mathbf{r}) \ne W_B(\mathbf{r})$, the momentum distribution given 
in Eq.~(\ref{eqn:nksc-def}) becomes
\begin{align}
n_{\rm CDW}(\mathbf{k}) = \frac{1}{M} \sum_{S,S^\prime} \sum_{\ell \in S, \ell' \in S'} & W_S^*(\mathbf{k})W_{S'}(\mathbf{k}) \nonumber \\
&\langle b_\ell^\dagger b_{\ell'} \rangle e^{-i\mathbf{k} \cdot (\mathbf{R_\ell}-\mathbf{R_{\ell'}})},
\end{align}
since the boson creation and annihilation operators have different weights
depending on their acting sublattice. Here the summation indices $\ell \in \{A, B\}$ 
and $\ell' \in \{A,B\}$ include the entire lattice.
This summation breaks up into terms that involve solely the $A$ sublattice, 
solely the $B$ sublattice, and terms that mix the $A$ and $B$
sublattices.  One can immediately see that the terms restricted to one 
of the sublattices are periodic with the periodicity of the reduced 
Brillouin zone, while the mixed terms are only periodic with respect to 
the full Brillouin zone. In general, these terms have different 
weightings when Wannier functions differ on two sublattices.
A detailed analysis of the CDW Hamiltonian given in Eq.~(\ref{eqn:cdwh}) 
and its momentum distribution is beyond the scope of this paper 
and they will be addressed elsewhere.

\end{document}